\documentclass{article}

\usepackage{xspace}
\newcommand{\framework}{Tahr\xspace}

\usepackage{url}
\usepackage{booktabs}
\newcommand{\remove}[1]{}
\usepackage{framed}
\usepackage[frozencache]{minted}
\usepackage{amsmath}
\usepackage{amsfonts}
\usepackage{algpseudocode}
\usepackage{algorithm}
\usepackage{algorithmicx}
\usepackage{booktabs}
\usepackage{listings}
\usepackage{enumitem}
\usepackage{tikz}
\usepackage{tcolorbox}
\usepackage{multicol}
\usepackage[margin=2.25cm]{geometry}

\usepackage{etoolbox,xpatch}
\makeatletter
\AtBeginEnvironment{minted}{\dontdofcolorbox}
\def\dontdofcolorbox{\renewcommand\fcolorbox[4][]{##4}}
\xpatchcmd{\inputminted}{\minted@fvset}{\minted@fvset\dontdofcolorbox}{}{}
\xpatchcmd{\mintinline}{\minted@fvset}{\minted@fvset\dontdofcolorbox}{}{} % see https://tex.stackexchange.com/a/401250/
\makeatother

\lstdefinestyle{myAG}{
    keywords={ \%\%, \%token, \%attribute, \%symbol, ->},
    sensitive=true, % keywords are not case-sensitive
    keywordstyle = \color{blue}
    morestring=[b]" % defines that strings are enclosed in double quotes
}
\begin{document}

\newcommand{\theTitle}{\framework: The Generative Attribute Grammar Framework}
\title{\theTitle}

\author{Matteo Ciccaglione, Pierciro Caliandro, Alessandro Pellegrini}

\maketitle

\begin{abstract}
In this article, we present \framework, a framework that allows taking attribute grammar specifications and generating a set of software artefacts that can be used programmatically to operate on text compliant with the grammars. \framework can be used as an algorithmic workbench to test different manipulations of attribute grammars and support translation between different languages out of the box. We describe the framework's organisation, how the user can specify an attribute grammar, and the generated software artefacts. We also discuss how \framework deals with ambiguous grammar specifications, and how this ambiguity can be effectively exploited when using attribute grammars for text generation. We test the correctness of \framework by showing the practical possibility of translating MIPS programs into their corresponding equivalents for x86 architectures and a custom virtual machine.
\end{abstract}

\section{Introduction}
\label{sec:intro}

Attribute grammars~\cite{Knuth1968-ed} represent a powerful formalism to specify semantic analyses associated with context-free grammars (CFGs), extending their descriptive capability beyond mere syntax toward semantics and computations inherent to programming languages and domain-specific languages. Fundamentally, attribute grammars enrich grammar productions with semantic rules and attributes, capturing dependencies among various elements of syntax trees through synthesised and inherited attributes. Synthesised attributes propagate semantic information from child nodes up to their parent nodes, while inherited attributes carry context-sensitive information downward or laterally across sibling nodes within a parse tree, enabling context-dependent interpretation and computations~\cite{Paakki1995-pp}.

The significance of attribute grammars arises from their capability to systematically address semantic correctness and generate efficient semantic analysers, facilitating compile-time checks such as type inference, consistency enforcement, and context-sensitive syntax validations. Moreover, they have been influential in compiler construction frameworks, driving the development of automatic compiler generation tools like yacc and ANTLR, which leverage concepts taken from attribute grammars to perform integrated semantic actions within syntactic parsing processes~\cite{Parr1995-vj,Aho2007-kw}, while not entirely supporting the definition of attribute grammars. Beyond traditional compiler design, more recent research highlights their utility in specifying domain-specific languages and model transformations, where semantics and constraints play a crucial role in expressing domain rules clearly and rigorously~\cite{Van-Wyk2007-rt}.

Attribute grammars enable formal and systematic facilitation of language translation tasks due to their explicit representation of semantic information and dependencies within syntax trees. By associating semantic rules with grammar productions, attribute grammars enable transformations of abstract syntax trees (ASTs), effectively supporting translation from a source language into a semantically equivalent representation in a target language. This feature has made attribute grammars particularly suitable for implementing compilers, transpilers, and translators between domain-specific languages, where attributes encode intermediate semantics used to generate accurate and context-sensitive translations that preserve the original program's correctness and intent~\cite{Paakki1995-pp,Van-Wyk2007-rt}.

Despite their foundational importance, practical adoption of attribute grammars has historically faced challenges, notably in performance, modularity, and scalability, driving contemporary research efforts towards optimisation strategies and efficient evaluation algorithms such as incremental and parallel attribute evaluation~\cite{Boyland2005-gn}. Recent developments have emphasised the extension of attribute grammars to address modular language specifications, incremental evaluation, and integration with modern model-driven engineering paradigms, ensuring their continued relevance and applicability in contemporary software engineering research~\cite{Erdweg2012-vf}. At the same time, we observe a lack of tools that efficiently support the definition of attribute grammars and generate tools capable of supporting the multifaceted tasks for which these grammars were devised, such as parsing, generating, and translating text. The gap is apparent if we compare this scenario with other forms of (possibly simpler) grammars, such as CFGs. Several tools exist for CFGs, such as yacc, Bison, ANTLR, or javacc, which allow transforming a grammar specification into a computer program able to parse text, verify its compliance with the specification, and generate structured representations of the text (often in the form of ASTs) that can be later manipulated or processed. However, such versatile tools do not exist for attribute grammars.

In this paper, we present \framework%
\footnote{Following the tradition of other grammar tools like yacc and Bison, we picked an animal-inspired name.},
the \textit{Translational Attribute Handler and
Rewriter}, a framework that allows taking attribute grammar specifications and generating a set of software artefacts that can be used programmatically to operate on text compliant with the grammars. \framework explicitly targets the multiple facets of attribute grammars, allowing to produce code that can \textit{parse} and \textit{generate} text compliant with the grammar specification. Furthermore, \framework allows to specify \textit{translation rules} that enable mapping one grammar into another. This feature, together with the capability to generate text that is compliant with a grammar specification, enables a simpler text translation from one language to another.

Moreover, \framework has been designed in a highly modular fashion, allowing its inner components and algorithms to be easily replaced. This organisation allows to rely on \framework also as an \textit{algorithmic workbench}, allowing users to test and experiment with different algorithms to manipulate grammars, both when parsing and generating text according to the grammar specification. We believe this is also an important contribution, exactly because attribute grammars have been observed to behave often poorly performance-wise. We consider a grammar workbench as an invaluable tool for conducting optimisation research in the context of attribute grammars.

Another motivation behind the development of \framework is to improve the protection of intellectual property (IP) of applications. Indeed, using \framework, we can transform one binary representation of an application on the fly into a semantically equivalent (but different) one. In this way, every execution of an application can result in a new generation of the code, making it harder for reverse engineering activities. This particular application of \framework is behind the scope of this paper, as it has already been discussed in previous work~\cite{Caliandro2025-xn}, yet without much focus on the used technique.

At the same time, in this paper we show the practical possibility of translating a program from one ISA to another. This is the third motivation of this work, namely supporting the translation of legacy applications, for which the original sources are lost or unavailable, to different, possibly more modern architectures. This is a critical software engineering goal, because there are many practical cases in which an existing application is extensively employed in business-critical scenarios, which anyhow force the users to rely on obsolescent hardware---for a discussion of these scenarios see, e.g.,~\cite{Bisbal1999-ay,Devereaux2010-ki,Seacord2003-bw}.

In \framework, we explicitly support ambiguous grammars by relying on non-determinism empowering stochasticity: if a rule is ambiguous when generating text, we randomly select one among the possible options. This design choice is essential for our IP protection use case. Indeed, a single portion of the program can be matched by multiple conflicting left-hand sides of the grammars: selecting one at random allows us to generate different ASTs from the same program, during parsing. Similarly, one rule can be associated with multiple productions: selecting one at random allows us to produce also different transformed versions from the very same AST. Overall, ambiguity managed through stochasticity enables us to provide an easier definition of \textit{translational attribute grammars}, and provides a powerful support for program manipulation.

When performing such manipulations, it is essential to preserve semantic equivalence. As mentioned, attribute grammars are exactly defined for this purpose, provided that the attributes' \textit{context} is manipulated correctly. For this purpose, we consider for each production a set of attributes that the user can explicitly manipulate when the production is applied. \framework provides a built-in support for set operations on attributes, but in the grammar specification the user can define their own operations in the form of functions, thus enabling any required form of attribute management.

The capability to inspect and manage attribute sets is also useful when dealing with ambiguous grammars and stochastocity: if a production rule is selected at random in a group of ambiguous rules, the generated program may detect some reasons to render the selection rule inapplicable, e.g. due to some user-defined constraint not being met while inspecting the attribute. In this case, \framework supports the dedicated \texttt{revert} statement, which instructs the program to abort the current production and select (randomly) a different one. We consider this strategy a simple method to realise complex grammars while preserving semantic equivalence during the translation.

We have assessed the viability and correctness of our approach by transforming a set of MIPS programs into their corresponding applications for x86 architectures and a custom virtual machine~\cite{Caliandro2025-xn}. By comparing the output of these programs, we can verify the correctness of non-trivial transformations in real-world applications. Moreover, this kind of transformation highlights the possibility of moving between CISC and RISC architectures, with very different capabilities, although generating programs that are not necessarily the most efficient ones. Anyhow, given the goal of supporting automatic translation for applications for which the original high-level sources are unavailable, these results can be considered perfectly acceptable.

We release \framework as open source---the source code associated with this paper is persistently available at \url{https://doi.org/10.5281/zenodo.15357268}, while the working tree can be found at \url{https://github.com/matteociccaglione/Tahr}---hoping that it will help the community to foster once again algorithmic research on attribute grammars and their translational capabilities, thanks to its workbench capabilities.

The remainder of this article is structured as follows. In Section~\ref{sec:relatex}, we discuss related work. Section~\ref{sec:background} presents a brief introduction to attribute grammar. In Section~\ref{sec:technical}, we present our framework along with the corresponding syntax defined for the input grammar. We outline an experimental assessment in Section \ref{sec:experimental}.

\section{Related Work}
\label{sec:relatex}

The literature on grammar-based parsing, semantic analysis, and automated compiler testing has developed a wide range of techniques, many of which pursue goals similar to those of \framework. However, they differ markedly in both their methodological assumptions and in the specific challenges they address.

A foundational contribution in this field is the work by Purdom~\cite{Purdom1972-fl}, which introduced an algorithm for generating a minimal set of test sentences from a context-free grammar. Each production rule is used at least once, allowing for systematic validation of parsers and grammar definitions. While Purdom's method remains purely syntactic and test-oriented, \framework builds upon a richer formalism: it integrates attribute grammars to support validation, semantic reasoning, and program translation. This shift from syntax to semantics allows \framework to operate on a broader spectrum of language manipulation tasks.

A different direction is explored by Moonen~\cite{Moonen2001-wt}, where \textit{island grammars} are proposed to handle irregular or incomplete input in reverse engineering scenarios. The core idea is to parse only the semantically meaningful code fragments (``islands''), ignoring unstructured or syntactically incorrect regions. In contrast, \framework assumes grammars that are fully specified at both the syntactic and semantic levels. Nevertheless, \framework acknowledges and accommodates ambiguity: its parser and generator components rely on stochastic resolution strategies when a rule is ambiguous. This probabilistic mechanism preserves grammatical rigour while allowing diversity in interpretation and output, a feature not present in island grammars' deterministic or error-tolerant models. The two approaches thus serve complementary purposes: Moonen's work emphasises fault tolerance in code analysis, whereas \framework focuses on controlled semantic transformations under formally defined rules.

Celentano et al.~\cite{Celentano1980-qe} describe a sentence generation system to support compiler testing. Their method starts from an extended BNF specification and iteratively generates semantically correct test programs, some of which may be intentionally incorrect to assess error handling. Although the validation and semantic correctness goals align with those of \framework, the two systems differ substantially. The Celentano approach focuses on coverage-driven generation and diagnostics, whereas \framework introduces a more expressive computational model that supports translation across language grammars and includes mechanisms for defining and executing semantic transformations within the grammar specification itself.

The classic work on orthogonal language description by van Wijngaarden~\cite{van-Wijngaarden1965-sp} marked a turning point in the formal treatment of programming languages. By introducing a meta-language capable of describing both syntax and semantics, it laid the groundwork for highly expressive but complex grammars, such as those used in ALGOL 68. \framework inherits some of these foundational ideas but applies them within a practical toolchain. It enables executable specifications of attribute grammars and implements these specifications through automatically generated libraries for parsing, code generation, and translation. In this sense, \framework realises in practice what van Wijngaarden's meta-language envisioned in theory.

The challenges and expressiveness of van Wijngaarden grammars are further explored by Augusto~\cite{Augusto2023-sc}, where he proposes restrictions that make these grammars computationally decidable. While van Wijngaarden grammars can express constructs beyond the Chomsky hierarchy, their undecidability often renders them impractical. \framework addresses this by adopting attribute grammars, which are powerful yet bounded enough to allow for effective computation. Using stochastic rule selection in \framework introduces non-determinism only in controlled contexts, such as the resolution of ambiguity, without sacrificing semantic determinism where it is essential.

Ghete~\cite{Ghete2018-ni} explicitly extends Bison to support attribute grammars, providing a practical mechanism for embedding semantic computations directly into Yacc-style grammar files. Although the underlying motivation resembles \framework's goals, the scope is much narrower. Ghete's system is confined to deterministic parser generation, with semantic actions integrated into a single monolithic tool. \framework, by contrast, defines a modular and extensible architecture that supports parsing, generation, and translation as separate but composable processes. Furthermore, \framework allows users to experiment with different algorithms and data structures, turning it into a research workbench for language manipulation. Such flexibility is not achievable within the more rigid structure of an extended Bison environment.

Unlike all these above proposals in the literature, \framework unifies syntactic structure and semantic evaluation in a single formalism, and supports non-deterministic yet semantically coherent translations between languages. In this way, \framework offers a framework that supports rigorous formal modelling and practical software generation.

Among the mature attribute grammar systems in the literature, Silver~\cite{Van-Wyk2010-vw} and JastAdd~\cite{Hedin2011-oy} represent the most extensively studied and widely adopted frameworks for extensible, analysable, and efficiently evaluable semantic specifications. Silver supports a modular, extensible variant of attribute grammars with static L-attributability checks and a runtime that constructs a dynamic dependency graph to allow efficient re-evaluation after local tree edits. JastAdd, by contrast, employs memoisation and circular attribute resolution strategies that accommodate arbitrary attribute dependencies while still enabling parallelism and lazy computation; its performance has been empirically validated on industrial-scale programs, with speed-ups exceeding an order of magnitude for specific dataflow analyses. In contrast, \framework sacrifices guaranteed determinism and strict static analyzability in favour of stochastic ambiguity resolution and translation diversity, aiming to support program transformation tasks rather than interactive editing or IDE support. While Silver and JastAdd emphasise correctness through static dependency constraints and runtime caching, our system leverages dynamic reduction and random selection to maximise syntactic expressiveness and translation variability, albeit without formal guarantees on confluence or incremental performance. As such, our work complements rather than competes with these systems, serving a distinct design space where translation flexibility and obfuscation potential outweigh the demands of statically guaranteed evaluation order.

Orthogonally, recent advances in large language models (LLMs) have significantly impacted the field of program synthesis and translation. Models such as Codex~\cite{Chen2021-ax}, CodeT5~\cite{Wang2021-oh}, and AlphaCode~\cite{Li2022-sx} have demonstrated the ability to generate syntactically valid and semantically meaningful code across multiple programming languages, often guided by natural language prompts or partial code fragments. These models are typically trained on vast corpora of real-world software, allowing them to capture idiomatic usage patterns and diverse translation strategies. However, despite their versatility, LLM-based approaches lack the formal guarantees and semantic transparency provided by grammar-based systems like \framework. In contrast to \framework's deterministic and interpretable translation pipeline---where user-defined rules explicitly govern every derivation step---LLM outputs are inherently opaque, non-reproducible, and difficult to verify. As such, LLMs are better suited for assistive or exploratory tasks, whereas \framework addresses correctness-critical scenarios requiring rigorous control over syntax, semantics, and code generation.

\section{Background on Attribute Grammars}
\label{sec:background}

Formally, a grammar $G$ is defined as a tuple $G = \langle N, T, P, S \rangle$, where $N$, $T$, $P$, and $S$ denote specific entities with well-defined roles~\cite{Chomsky2019-nh}. Specifically, the set $N$ represents non-terminal symbols, each corresponding to syntactic categories or intermediate constructs used in defining the grammar structure. The set $T$ consists of terminal symbols, which constitute the actual symbols appearing in the language generated by the grammar, forming its alphabet. Notably, these sets must satisfy the condition $N \cap T = \emptyset$. The set $P$ represents productions or rules, each defined as a pair $(\alpha, \beta)$, commonly expressed as $\alpha \rightarrow \beta$, with $\alpha \in (N \cup T)^*\cdot N\cdot (N \cup T)^*$ and $\beta \in (N \cup T)^*$. Finally, $S\in N$ identifies the start symbol, indicating the initial symbol from which derivations of strings in the language begin.  

Grammars can be categorised in \textit{prescriptive}, \textit{descriptive}, and \textit{generative}, representing distinct conceptualisations of grammars, each differing in objective and application~\cite{Chomsky1965-nf}. Prescriptive grammars impose normative standards, defining how language should be used correctly according to particular linguistic authorities, hence serving as guidelines to enforce correctness and stylistic consistency. In contrast, descriptive grammars document actual linguistic usage without prescribing rules, focusing empirically on how native speakers naturally employ language. They record and analyse observable patterns, variations, and norms in speech and writing rather than enforcing compliance with idealised rules. Generative grammars, notably advanced by Noam Chomsky, differ fundamentally from both previous types by providing formalised systems capable of explicitly generating all and only the grammatically valid sentences of a language, thereby focusing on underlying syntactic structures and competence rather than observed performance. Generative grammars thus represent a theoretical framework for understanding linguistic universals and internal cognitive processes involved in language acquisition and usage.

Formally, an attribute grammar can be defined as a tuple $AG = (G, A, R, C)$, where each component is structured as follows. The underlying context-free grammar is $G = (N, T, P, S)$, with $N$ denoting the set of non-terminal symbols, $T$ the set of terminal symbols, $P$ the set of productions, and $S$ the start symbol of the grammar. The set $A$ specifies the attributes, which are partitioned for each grammar symbol $X \in N \cup T $ into two disjoint subsets: inherited attributes and synthesised attributes. Usually, only synthesised attributes are defined for terminal symbols, whereas non-terminal symbols may possess both inherited and synthesised attributes.

The set $R$ comprises the semantic rules, each associated with a grammar production $p \in P$. Formally, for a given production of the form:

\begin{equation}
X_0 \rightarrow X_1 X_2 \dots X_n,\quad \text{with } X_i \in N \cup T,\quad 0 \leq i \leq n    
\end{equation}

Semantic rules are equations defined in terms of attribute occurrences associated with symbols in the production, such as:

\begin{equation}
    X_i.a := f(X_{j_1}.a_1, X_{j_2}.a_2, \dots, X_{j_k}.a_k)
\end{equation}

where $X_i.a$ denotes the attribute $a$ of the grammar symbol $X_i$, and $f$ represents an attribute evaluation function.

Finally, $C$ is the set of attribute constraints or conditions, typically predicates expressed over attributes to enforce semantic consistency or correctness.

Attribute grammars belong to the broader class of generative grammars, specifically extending the formalism of CFGs with semantic attributes and evaluation rules to incorporate semantic analysis directly into syntactic descriptions. The reason for this classification lies primarily in their explicit generative capacity: attribute grammars generate syntactic structures (ASTs) coupled with attribute values that propagate semantic information. Unlike descriptive grammars, which record observed usage, or prescriptive grammars, which enforce correctness rules, attribute grammars formally define semantics by systematically assigning meaning through attribute computations attached to grammar productions. By integrating syntax and semantics into a unified formalism, attribute grammars provide a generative mechanism powerful enough to derive syntactically correct strings and to evaluate context-dependent semantic properties, such as type checking, expression evaluation, and translation tasks, making them a robust framework within generative language theory.

\section{TAHR}
\label{sec:technical}

\framework is a comprehensive formal system designed to facilitate the specification, parsing, generation, and translation of languages and structured textual artefacts through attribute grammars enriched with explicit semantic and translational rules. Conceptually based on the programmability of parser generators such as Bison~\cite{Donnelly2015-rr}, this framework provides users with constructs to describe grammatical structures and embed attribute-based semantic evaluations directly into grammar productions.

\begin{figure}[t]
    \centering
    \includegraphics[width=0.6\linewidth]{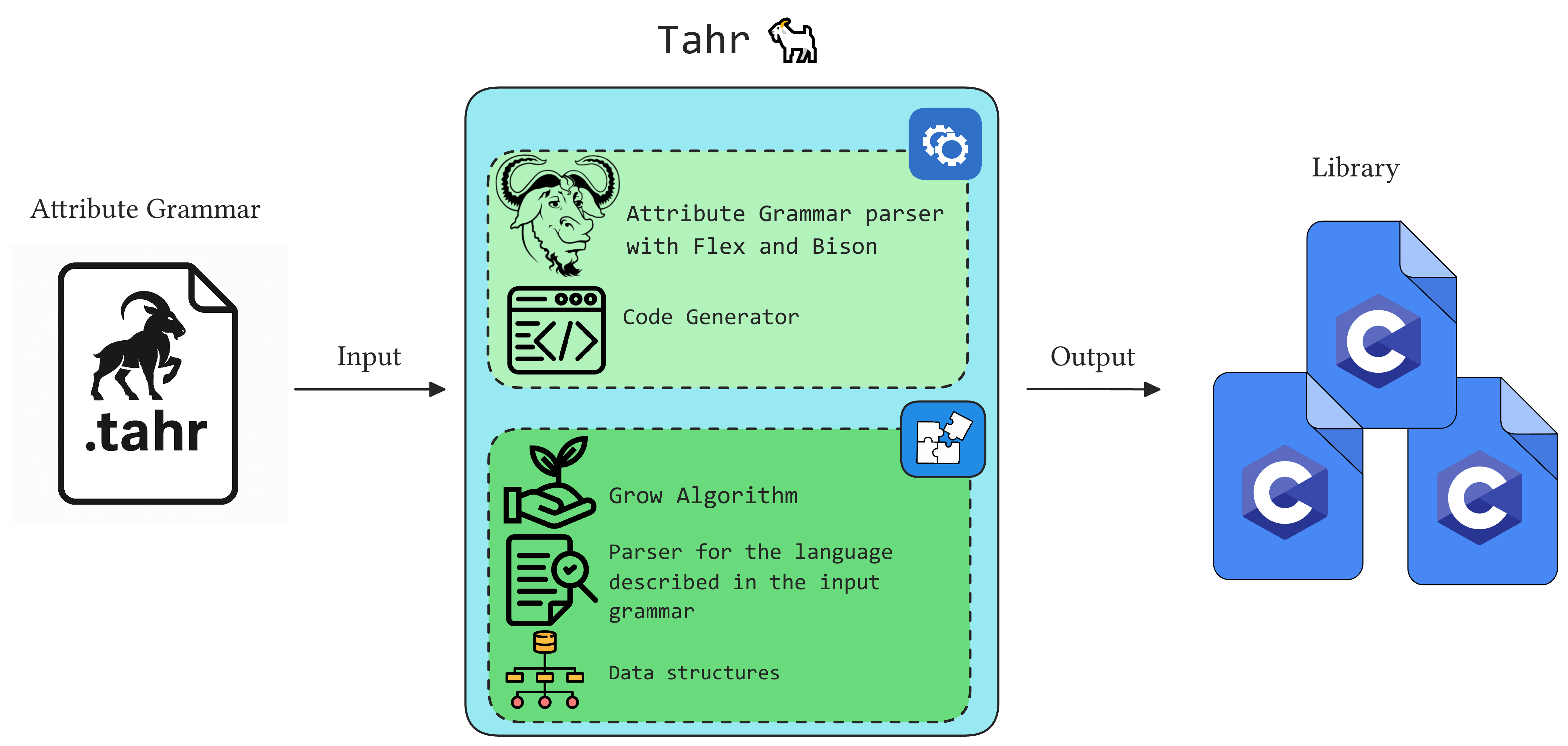}
    \caption{Overall Structure of the \framework framework.}
    \label{fig:framework-struct}
\end{figure}

A key feature of \framework is its expressive translational capability, where grammar(s) specifications can be explicitly combined via translational rules. Such rules systematically map semantic information encoded in attributes from one grammar specification to corresponding semantic structures in another, enabling structured text transformations from source to target languages. Importantly, these transformations are not necessarily deterministic; rather, the framework supports encoding stochastic rules and programmatic interventions during attribute evaluation, permitting selective interruption and dynamic redirection of transformations based upon runtime semantic checks or probabilistic conditions. In this manner, \framework serves as an advanced computational model where semantic actions, constraints, and stochastic conditions are explicitly encoded into attribute computations, thus bridging declarative grammar specifications and imperative software behaviour in a structured, formalised manner.

The modular organisation of \framework is depicted in  Figure~\ref{fig:framework-struct}. As we mentioned, we also designed it to serve as an attribute grammar workbench so that different components, algorithms, and data structures can be plugged in for experimentation and benchmarking purposes. Following the strategy of other tools such as flex, yacc, and Bison, the user of the framework is expected to provide an \textit{extended attribute grammar} specification, which is processed to transform it into C sources that implement the various parsing, generation, and translation tools. These sources can then be linked into a destination program or compiled into a (shared) library to export their functionality to other applications. Internally, we rely on Flex and Bison to interpret the specified attribute grammar, but the framework provides (pluggable) algorithms and data structures that can be modified or replaced to fine-tune the generated output sources.

In the remainder of this Section, we first discuss the specification of the syntax of the input file to \framework, which allows the user to specify an extended attribute grammar that can be used for parsing, generating, and translating concepts. Then, we discuss the algorithms and data structures shipped by default, along with the internal API that allows the workbench capabilities to be exploited. Finally, we discuss the organisation of the generated output sources.

\begin{listing}[t]
\begin{multicols}{2}
\begin{minted}{asm}
    fibonacci:
        daddiu  $sp, $sp, -36
        sd      $ra, 32($sp)
        sw      $s1, 20($sp)
        lw      $s0, 20($sp)
        slti    $s0, $s0, 2             
        beqc    $s0, $zero, BB0_3
        bc      BB0_2
    BB0_2:
        lw      $s0, 20($sp)
        sw      $s0, 24($sp)
        bc      BB0_4
    BB0_3:
        lw      $s0, 20($sp)
        addiu   $s1, $s0, -1
        balc    fibonacci
        sw      $v0, 16($sp)
        lw      $s0, 20($sp)
        addiu   $s1, $s0, -2
        balc     fibonacci
        move    $s0, $v0
        lw      $v0, 16($sp)
        addu    $s0, $v0, $s0
        sw      $s0, 24($sp)
        bc       BB0_4
    BB0_4:
        lw      $v0, 24($sp)
        ld      $ra, 32($sp)
        daddiu  $sp, $sp, 36
        jrc     $ra
\end{minted}
\end{multicols}
\caption{Reference Example Program: Fibonacci in MIPS Assembly.}
\label{lst:fibonacci}
\end{listing}

To illustrate how the user can employ \framework and describe its inner workings, we rely on the example program in Listing~\ref{lst:fibonacci}. This program computes the $n$-th Fibonacci number (with $n$ specified in input) and is implemented in MIPS assembly. We will show how it is possible to encode in \framework an attribute grammar to parse the code in Listing~\ref{lst:fibonacci} and generate a working equivalent in x86 assembly. For this purpose, we will define translational rules between the two parts of the grammar specification and apply a (limited) amount of stochastic generation.  Stochasticity is not entirely necessary for this example, but allows us to illustrate this particular feature of \framework.

\subsection{Attribute Grammar(s) Specification}
\label{subsec:grammar}

\begin{figure}[t]
    \centering
    \includegraphics[width=.6\linewidth]{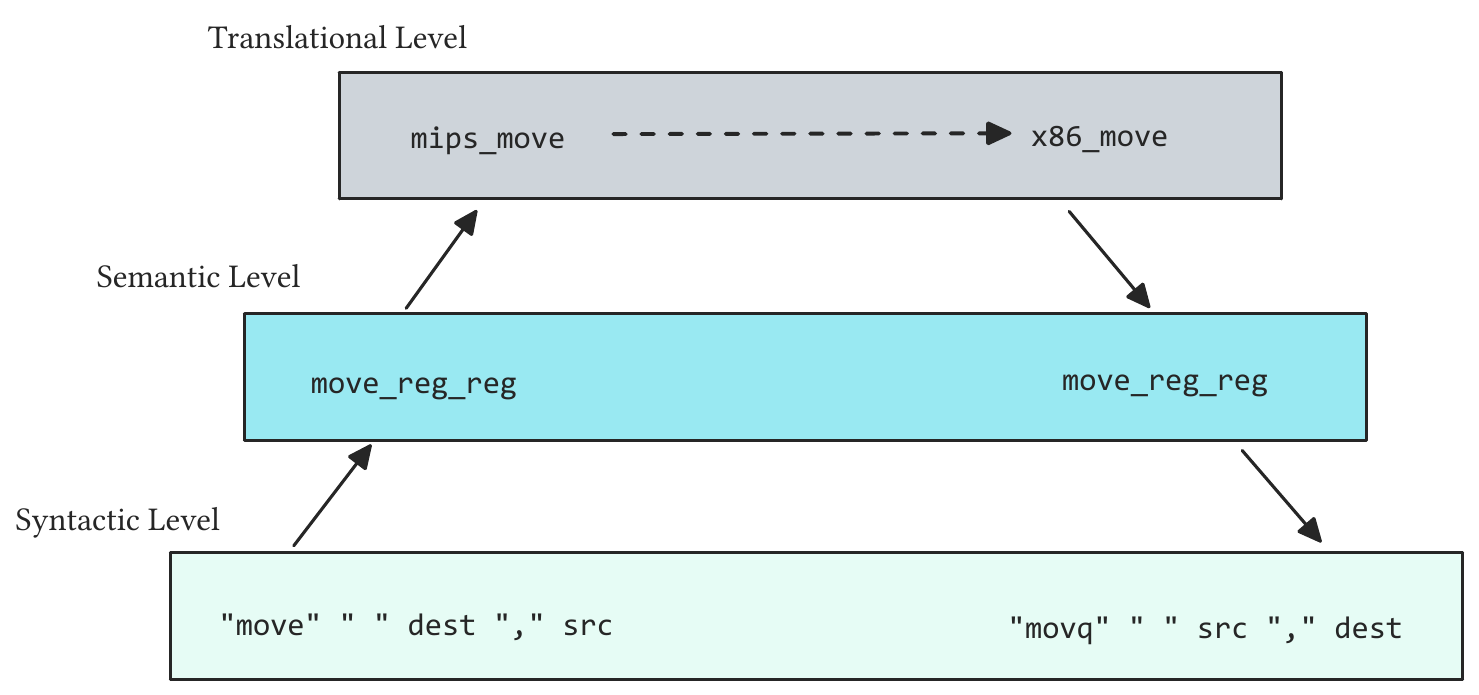}
    \caption{\framework 3-level extended attribute grammar.}
    \label{fig:grammar-levels}
\end{figure}

Overall, the formal description of linguistic structures in a \framework extended grammar is organised into a hierarchy composed of three interdependent levels, as depicted in Figure~\ref{fig:grammar-levels}.  We also highlight the dependencies among the levels with arrows and how we can use rules to move across the different levels.

The \textit{syntactic level} at the bottom defines the formal structures of expressions by means of production rules, determining grammatical hierarchies and relationships necessary to ensure structural correctness of sentences.  Above, the \textit{semantic level} assigns meanings through logical operations and computations on attributes associated with syntactic structures; these attributes, propagated across the syntactic tree, allow correct interpretation, evaluation, and manipulation of linguistic expressions.  Finally, at the highest position, the \textit{translational level} plays a fundamental role as it provides a formal and abstract representation, enabling translation between distinct grammars.

This translational level serves as a logical interface between different linguistic systems, systematically mapping structures from a source grammar into corresponding structures of a target grammar. Thus, the translational grammar governs the translation process, facilitating accurate and consistent formal conversions as required in the extended grammar specification~\cite{Hutchins1992-bu,Baker2018-da}.

Beyond this logical organisation of the grammar, we recall that we can also include in the specification a \textit{programmatic} part, in a way similar in spirit to Bison grammars, to support, e.g., the definition of operations on attribute sets in production rules. To support all these different aspects of the grammar specification, a grammar file is sectioned into the following several \textit{parts}, namely a \textit{header}, a \textit{declaration} part (level 1), a \textit{representation} part (still level 2), an \textit{expansion} part (level 2) and a \textit{translation} part (level 3).
The token \texttt{\%\%} is used as a part separator.

The \textit{header} can be used to extend the generated library by including additional header (.h) files, thus enabling the user to rely on an external library of any form, for greater generality in the development of attribute manipulation functions. These header files must be enclosed in a \texttt{\%\{\}\%} block. The header can also include the definition of start conditions, similar to those that can be defined in Flex~\cite{Paxson2023-ey}~\cite{Levine2009-dl}, useful for preventing potential parsing conflicts. For our Fibonacci example, the header part is quite simple, as reported in Figure~\ref{lst:example-header}: we only include a C header file, which exposes functions later used for register mapping across the different architectures (see below).

\begin{listing}[!ht]
\begin{minted}{asm}
    %{
            #include "utils.h"
    %}
\end{minted}
\caption{Example header to translate Listing~\ref{lst:fibonacci}.}
\label{lst:example-header}
\end{listing}

The \textit{declaration} part is used to define symbols (terminal and non-terminal) and attributes, along with their type/values. Different keywords can be used to do this, namely:

\begin{itemize}[noitemsep]
    \item \texttt{\%token}: It is used to define a set of terminal symbols, indicated immediately after the keyword and separated by commas;
    \item \texttt{\%attribute}: It is used to define an attribute that can be used in grammar. An attribute is represented by: a name, a data type (chosen from those in the table) and a set of values it can assume, or a regular expression in the case of infinite values (optional). The syntax requires that the keyword be followed by a comma-separated string organised as follows: attribute name, type, values, for example: \texttt{\%attribute test1, char*,"t1","t2","t3"}. Regular expressions can be used, if enclosed within a pair of \texttt{!!} characters. Additionally, the user can include code to manipulate the current start condition using the BEGIN statement. This code should be enclosed in curly braces, similar to attribute operations (see the expansion section).
    \item \texttt{\%symbol}: It is used to define a non-terminal symbol, represented by: a name and a set of attributes (optional). Since a symbol can use multiple instances of the same attribute, it is possible to assign a name to each attribute, along with its type (which must have been previously defined by an \%attribute statement). For example \texttt{\%symbol sym,test1,myatt1,test1,myatt2}.
\end{itemize}

\begin{listing}[!t]
\begin{minted}{asm}
    //Mips part
    %token LOAD_DW, LOAD_W, STORE_D, STORE_W, S_L_R_I, BRANCH_EQ_C, BRANCH_COM, JUMP_R_C, BALC, CODE_L, ADDU_R_I, MOVE, ADDU_R_R, DADDU_R_I;
    %attribute reg, char*, "$zero", "$v0", "$s0", "$s1", "$sp", "$ra";
    %attribute num, int;
    %attribute label, char*, !![\.]*[a-zA-Z0-9_$]+[:]*!!;
    
    // MIPS Isa
    %symbol addu_r_r, reg, reg1, reg, reg2, reg, dest;
    %symbol addu_r_i, reg, reg1, num, imm, reg, dest;
    %symbol daddu_r_i, reg, reg1, num, imm, reg, dest;
    %symbol load_dw, reg, dest, num, offset, reg, base;
    %symbol load_w, reg, dest, num, offset, reg, base;
    %symbol store_d, reg, source, num, offset, reg, base;
    %symbol store_w, reg, source, num, offset, reg, base;
    %symbol move, reg, dest, reg, src;
    %symbol set_l_r_i, reg, reg1, num, imm, reg, dest;
    %symbol code_label, label, text;
    %symbol balc, label, target;
    %symbol branch_eq_c, reg, reg1, reg, reg2, label, target;
    %symbol jump_r_c, reg, reg1;
    %symbol branch_c, label, target;
    
    // x86 part
    %token I_ADD_R_R, I_ADD_I_R, I_CALL, I_CMP_R_R, I_CMP_R_I, I_JB, I_JMP, I_MOV_R_R, I_MOV_M_R, I_MOV_R_M, I_JZ, I_MOV_I_R, I_RET;
    %attribute ireg, char*, "%rsp", "%r13", "%edi", "%esi", "%r12d", "%edx", "%rdx", "%rsi";
    %attribute suf, char*, "q", "l", "w", "b";
    %symbol i_ret;
    %symbol i_add_r_r, suf, suffix, ireg, reg1, ireg, dest;
    %symbol i_add_i_r, suf, suffix, num, imm, ireg, dest;
    %symbol i_call, label, text;
    %symbol i_cmp_r_r, suf, suffix, ireg, reg1, ireg, reg2;
    %symbol i_cmp_r_i, suf, suffix, ireg, reg1, num, imm;
    %symbol i_jb, label, text;
    %symbol i_jz, label, text;
    %symbol i_jmp, label, text;
    %symbol i_mov_i_r, suf, suffix, ireg, dest, num, imm;
    %symbol i_mov_r_r, suf, suffix, ireg, reg1, ireg, dest;
    %symbol i_mov_m_r, suf, suffix, num, offset, ireg, base, ireg, dest;
    %symbol i_mov_r_m, suf, suffix, num, offset, ireg, base, ireg, src;
    
    // Intermediate symbols needed for translation
    %symbol add_r_i_t, ireg, reg1, num, imm, ireg, dest, suf, suffix;
    %symbol add_r_r_t, ireg, reg1, ireg, reg2, ireg, dest, suf, suffix;
    %symbol set_l_r_i_t, ireg, reg1, num, imm, ireg, dest, suf, suffix; 
    %symbol branch_eq_r_r_t, ireg, reg1, label, address, ireg, reg2, suf, suffix;
\end{minted}
\caption{Example declaration part to translate Listing~\ref{lst:fibonacci}.}
\label{lst:example-declaration}
\end{listing}

In Listing~\ref{lst:example-declaration}, we report the declaration part of our MIPS-to-x86 grammar used to translate the Fibonacci example. The listing shows how the grammar formalism is used to explicitly declare the necessary terminal symbols corresponding to MIPS instructions, such as \texttt{li}, \texttt{move}, \texttt{add}, and control flow primitives like \texttt{bge} and \texttt{j}, all grouped under the \texttt{\%token} directive. The attributes of such instructions, such as \texttt{label}, \texttt{reg}, and \texttt{imm}, are declared using the \texttt{\%attribute} keyword to support lexemes representing labels, registers, and immediate values, respectively, with types \texttt{char *} or \texttt{int}.
Some attributes are defined over finite domains by enumeration, while others are specified through regular expressions enclosed within \texttt{!!} delimiters to capture infinite classes of lexemes. 

The listing also contains a set of non-terminal symbol declarations using the \texttt{\%symbol} keyword. Each \texttt{\%symbol} line defines a non-terminal symbol along with its associated typed attributes. These attributes, previously declared using \texttt{\%attribute}, are attached to the symbol with local names, allowing multiple instances of the same attribute type to coexist within a single symbol. For example, the declaration \texttt{\%symbol addu\_r\_r,reg,reg1,reg,reg2,reg,dest} introduces a non-terminal for the MIPS instruction \texttt{addu} with three register attributes named \texttt{reg1}, \texttt{reg2}, and \texttt{dest}, capturing the structure of an \texttt{addu} MIPS instruction and defining the number of attributes used to specify the semantics of their instances. The same description for the x86 part of the grammar is provided.

In this part of the grammar, we have also introduced intermediate non-terminal symbols that bridge the MIPS and x86 representations. These symbols, such as \texttt{add\_r\_i\_t} or \texttt{set\_l\_r\_i\_t}, include attributes from both architectures. These translation-specific symbols support rewriting rules that convert abstract MIPS instructions into abstract x86 forms, allowing merging of the parsing and the generation features of attribute grammars into a single grammar specification. In this sense, we could see a \framework grammar specification as a combination of multiple grammars that can coexist thanks to these intermediate symbols (and rules, as we will show later), according to the translational scheme in Figure~\ref{fig:grammar-levels}. 

The \textit{representation} section defines how terminal symbols are translated into text. Specifically, it enables the construction of meaningful sentences by combining the various attributes of a symbol. To achieve this, we define a representation rule---a rule that expands a terminal symbol into a string formatted as follows: elements enclosed in quotation marks are treated as constants, while all other elements must correspond to attributes of the equivalent non-terminal symbol. Elements must be separated by one or more spaces. The rule format uses a colon (\texttt{:}) to separate the left-hand side (LHS), which contains the terminal symbol, from the right-hand side (RHS), which contains the structured string.

\begin{listing}[t]
\begin{minted}{asm}
    // Mips
    JUMP_R_C: "jrc" " " reg1 ;
    BRANCH_EQZ_C: "beqzc" " " reg1 "," target ;
    BRANCH_COM: "bc" " " target ;
    BALC: "balc" " " target ;
    DADDU_R_I: "daddiu" " " dest "," reg1 "," imm ;
    ADDU_R_R: "addu" " " dest "," reg1 "," reg2 ;
    MOVE: "move" " " dest "," src ;
    CODE_L: text ;
    LOAD_DW: "ld" " " dest "," offset "(" base ")" ;
    LOAD_W: "lw" " " dest "," offset "(" base ")" ;
    STORE_D: "sd" " " source "," offset "(" base ")" ;
    STORE_W: "sw" " " source "," offset "(" base ")" ;
    S_L_R_I: "slti" " " dest "," reg1 "," imm ;
    
    // x86
    I_RET: "ret" ;
    I_JZ: "jz" " " text ;
    I_ADD_R_R: "add" suffix " " reg1 "," dest ;
    I_ADD_I_R: "add" suffix " " "$" imm "," dest ;
    I_MOV_I_R: "mov" suffix " " "$" imm "," dest ;
    I_CALL: "call" " " text ;
    I_CMP_R_R: "cmp" suffix " " reg1 "," reg2 ;
    I_CMP_R_I: "cmp" suffix " " "$" imm "," reg1 ;
    I_JB: "jl" " " text ;
    I_JMP: "jmp" " " text ;
    I_MOV_R_R: "mov" suffix " " reg1 "," dest ;
    I_MOV_M_R: "mov" suffix " " offset "(" base ")" "," dest ;
    I_MOV_R_M: "mov" suffix " " src "," offset "(" base ")" ; 
\end{minted}
\caption{Example representation part to translate Listing~\ref{lst:fibonacci}.}
\label{lst:example-representation}
\end{listing}

An example for our MIPS-to-x86 translation is shown in Listing~\ref{lst:example-representation}. The listing contains a series of representation rules that map terminal symbols to concrete assembly strings by composing constant fragments with named attributes drawn from the corresponding non-terminal declarations. Each rule follows the formal structure introduced above: the LHS is a terminal symbol, and the RHS is a sequence of constants and placeholders. Constants are enclosed in quotes and reproduced verbatim in the final output or matched exactly when parsing; unquoted elements correspond to attribute names and are resolved at runtime based on the values assigned in the semantic model.

The first part of the listing encodes the representation of MIPS instructions. For example, the rule for \texttt{DADDU\_R\_I} produces an output like \texttt{daddiu dest, reg1, imm}, where \texttt{dest}, \texttt{reg1}, and \texttt{imm} are mapped to the actual values of the corresponding attributes. Addressing modes are also handled explicitly, as seen in \texttt{LOAD\_DW}, which translates into a string like \texttt{ld dest, offset(base)}, closely resembling the MIPS syntax. Labels are handled as raw textual entries in rules such as \texttt{CODE\_LABEL: text}, directly corresponding to the label name without quoting or formatting.

The second half of the listing covers x86 instruction formats, where the conventions of AT\&T assembly syntax are respected.  nstructions like \texttt{I\_ADD\_I\_R} and \texttt{I\_MOV\_I\_R} prepend the immediate operand with a dollar sign to comply with the expected syntax form (e.g., \texttt{addq \$imm, \%reg}). Suffixes are handled dynamically via the suffix attribute to reflect operand sizes (\texttt{q}, \texttt{l}, etc.), enabling a uniform pattern across variants of the same instruction. Memory operations, such as \texttt{I\_MOV\_M\_R}, demonstrate how structured addressing is reproduced by interpolating the offset, base, and other operands into the familiar format offset(base), following System V and AT\&T conventions.

Therefore, this part of the \framework grammar specification provides the textual backbone for the code translation process that we are aiming at, ensuring that each symbol in the intermediate model is grounded in a concrete syntactic structure tailored to the target architecture (either MIPS or x86).

The \textit{expansion} section defines the derivation rules through which non-terminal symbols are decomposed into sequences of other non-terminal or terminal symbols, specifying the recursive structure of the language. Each expansion rule comprises an LHS, which must be a single non-terminal symbol declared in the grammar, and an RHS, which consists of one or more grammar symbols arranged in a specific order. The two sides are separated by the arrow symbol (\texttt{->}), which denotes the syntactic derivation. The syntax supports using the pipe (\texttt{|}) to separate multiple RHS variants within a single rule, allowing the specification of alternative expansions of the same non-terminal. This mechanism generalises the pattern-matching capabilities introduced in the representation section and significantly increases the expressive power of the specification, enabling the definition of complex hierarchical and recursive constructs essential for syntactic analysis and semantic transformation.

In addition to specifying the structural composition of symbols, the \textit{expansion} section also supports the definition of attribute operations, which enable the dynamic computation and propagation of semantic information during rule application. These operations are expressed as embedded C code fragments executed whenever a rule is matched. Attribute access within these fragments follows this convention: attributes of the LHS non-terminal are accessed via the \texttt{\$\$} notation, while attributes of the RHS symbols are accessed using \texttt{\$n}, where \texttt{n} is the 1-based index of the corresponding symbol in the RHS sequence---for example, \texttt{\$1} refers to the attributes of the first symbol in the RHS. This design gives the grammar specification full control over how attributes are computed and assigned.

To support enumeration or symbolic attribute value binding, the grammar allows accessing the set of predefined attribute values through automatically generated arrays named \texttt{\{att\_name\}\_values}, where \texttt{\{att\_name\}} is the identifier of the attribute as declared in the \texttt{\%attribute} section. These arrays can be accessed from the embedded code to assign values from the declared domain. This mechanism permits the use of high-level semantic abstractions without constraining the expressiveness of the underlying translation logic. Overall, attribute operations form the semantic foundation of the grammar, supporting the manipulation of symbol instances and aiding in the creation of detailed intermediate representations customised for the requirements of the target language or execution model.

The expansion mechanism also supports conditional constraints on attribute values, enabling fine-grained control over rule applicability by using the \texttt{if} construct, following standard C syntax. Within the body of the attribute operation, it is possible to express conditions that must be satisfied for the transformation to proceed. If such a condition fails, the special \texttt{revert} keyword can be invoked to signal that the current rule application should be abandoned. When generating the final program that can process a \framework grammar, the \texttt{revert} directive is expanded into dedicated code that restores the internal state to what it was before the rule's application, ensuring semantic consistency and allowing alternative derivations to be considered.

Importantly, the invocation of \texttt{revert} does not simply abort the current expansion path---it also informs the probabilistic selection mechanism associated with the grammar engine. When a rule is reverted, the system adjusts the likelihood of that branch being selected again under similar conditions, effectively introducing a dynamic bias into the stochastic generation process. This feedback mechanism makes it possible to steer the derivation process away from semantically invalid or undesirable configurations without statically disabling the rule, thereby preserving the generative power of the grammar while embedding runtime semantic filtering.

\begin{listing}[!t]
\begin{multicols}{2}
\begin{minted}{asm}
    //Mips
    //Non-determinism
    i_push_r -> i_add_i_r,i_mov_r_m{
        $1.suffix = $$.suffix;
        $2.suffix = $$.suffix;
        $1.imm = -8;
        $1.dest = ireg_values[4]; //rsp
        $2.src = $$.reg1;
        $2.base = ireg_values[4];
        $2.offset = 0;
    }| I_PUSH_R;
    i_pop_r -> i_mov_m_r,i_add_i_r{
        $1.suffix = $$.suffix;
        $2.suffix = $$.suffix;
        $2.imm = 8;
        $2.dest = ireg_values[4];
        $1.dest = $$.reg1;
        $1.base = ireg_values[4];
        $1.offset = 0;
    }| I_POP_R;
    i_mov_r_r -> i_push_r,i_pop_r{
        //Only for movq instruction
        if ($$.suffix != suf_values[0]){
                revert; 
        }
        $1.suffix = suf_values[0];
        $2.suffix = suf_values[0];
        $1.reg1 = $$.reg1;
        $2.reg1 = $$.dest;
    } | I_MOV_R_R;
    ...
    //x86
    i_ret -> I_RET;
    ...
    //From translated symbol to x86 instruction
    add_r_r_t -> i_mov_r_r,i_load_mem,i_add_r_r{
        if ($$.dest==$$.reg1 || $$.dest==$$.reg2) {
            revert;
        }
        $1.suffix = $$.suffix;
        $3.suffix = $$.suffix;
        $1.dest = $$.dest;
        $1.reg1 = $$.reg1;
        $2.reg1 = $$.reg2;
        $2.old1 = $$.reg2;
        $3.dest = $$.dest;
        $3.reg1 = $$.reg2;
    } | i_add_r_r{
        $1.suffix = $$.suffix;
        if ($$.dest == $$.reg2) {
            $1.dest = $$.dest;
            $1.reg1 = $$.reg1;
        }
        else if ($$.dest == $$.reg1) {
            $1.dest = $$.dest;
            $1.reg1 = $$.reg2;
        }
        if ($$.dest != $$.reg1 && $$.dest != $$.reg2) {
            revert;
        }
    };
    add_r_i_t -> i_mov_r_r,i_add_i_r {
        if ($$.dest == $$.reg1) {
            revert;
        }
        $1.suffix = $$.suffix;
        $2.suffix = $$.suffix;
        $1.dest = $$.dest;
        $1.reg1 = $$.reg1;
        $2.dest = $$.dest;
        $2.imm = $$.imm;
    } | i_add_i_r{
        $2.suffix = $$.suffix;
        if ($$.dest != $$.reg1) {
            revert;
        }
        $1.dest = $$.dest;
        $1.imm = $$.imm;
    };
    branch_eq_r_r_t -> i_cmp_r_r,i_jz {
        $1.suffix = $$.suffix;
        $1.reg1 = $$.reg1;
        $1.reg2 = $$.reg2;
        $2.text = $$.address;
    };
    set_l_r_i_t -> i_cmp_r_i,i_jb,i_mov_i_r,i_jmp,code_label,i_mov_i_r,code_label {
        $1.suffix = $$.suffix;
        $3.suffix = $$.suffix;
        $6.suffix = $$.suffix;
        $1.reg1 = $$.reg1;
        $1.imm = $$.imm;
        $2.text = create_label();
        $3.imm = 0;
        $3.dest = $$.reg2;
        $4.text = create_label();
        $5.text = print_as_a_label($2.text);
        $6.dest = $$.reg2;
        $6.imm = 1;
        $7.text = print_as_a_label($4.text);
    };
    branch_lt_c_t -> i_cmp_r_r,i_jb {
        $1.reg2 = $$.reg1;
        $1.reg1 = $$.reg2;
        $1.suffix = suf_values[1];
        $2.text = $$.target;
    };
\end{minted}
\end{multicols}
\caption{Example expansion part to translate Listing~\ref{lst:fibonacci}.}
\label{lst:example-expansion}
\end{listing}

For our MIPS-to-x86 grammar, an excerpt of the expansion part is provided in Listing~\ref{lst:example-expansion}. This section illustrates how \framework supports non-deterministic expansion through multiple alternative right-hand sides, attribute-based control logic, and structured sequencing of translated operations. The listing includes compound expansion rules for both translated and low-level x86 constructs. For instance, the \texttt{i\_push\_r} symbol may be expanded either into a sequence of \texttt{i\_add\_i\_r} and \texttt{i\_mov\_r\_m} instructions or directly into a terminal rule \texttt{I\_PUSH\_R}. Attribute operations assign fixed offsets and determine the register locations involved in emulating the push operation. Similarly, the \texttt{i\_pop\_r} symbol conditionally expands to a combination of memory load and stack pointer adjustment instructions, preserving the correct calling convention.

Translated symbols like \texttt{add\_r\_r\_t} and \texttt{add\_r\_i\_t} further demonstrate the flexibility of this mechanism. Both symbols expand to sequences of instructions that may involve register copying, arithmetic, and memory operations, depending on the register allocation and operand overlap. Conditional logic using revert is employed to guarantee the correctness of register use---ensuring, for example, that no instruction overwrites a source register that is still in use.

Control-flow constructs are also handled declaratively. The \texttt{branch\_eq\_r\_r\_t} symbol expands into a \texttt{cmp} followed by a conditional jump \texttt{jz}, and more complex constructs like \texttt{set\_l\_r\_i\_t} expand into multi-instruction sequences involving comparisons, conditional and unconditional jumps, and label management. The generation of new labels via \texttt{create\_label()} and their string representation using \texttt{print\_as\_a\_label()} shows the possibility to rely on custom C functions when writing a \framework grammar.

The \textit{translation} section extends the expressive capabilities of the grammar by enabling its use as a model for language-to-language transformation. This feature permits the specification of at most two distinct languages within a single grammar and defines explicit translation correspondences between their non-terminal symbols. Through translation rules, it becomes possible to map constructs from the source language to equivalent or semantically related constructs in the target language, thus enabling a generative approach to translation that integrates with the expansion mechanism. The overall effect is the specification of a non-deterministic translator, where translation steps are interleaved with structural derivation.

Syntactically, translation rules resemble expansion rules but differ in several key aspects. Rach rule consists of an LHS and an RHS separated by a two-headed arrow \texttt{<->}, indicating a symmetric association between two non-terminal symbols. The RHS must contain exactly one symbol, unlike the expansion syntax that allows for arbitrary-length RHS sequences. Furthermore, each non-terminal symbol may appear at most once as the LHS in the set of translation rules, enforcing a functional view of translation where a given source construct has a unique primary mapping to its target counterpart.

Attribute operations in translation rules follow the same conventions as in expansion rules. Attributes of the LHS symbol are accessed through the \texttt{\$\$} notation, while attributes of the RHS symbol are accessed using enumerated symbols like \texttt{\$1}. Using these operations, it is possible to assign values, propagate attributes, or compute derived values using C code, including references to attribute value arrays, where appropriate. 

Unlike expansion rules, the use of the \texttt{revert} keyword is explicitly disallowed in translation rules. This restriction reflects the semantic interpretation of translation as a committed mapping step: once a translation rule is selected, it must deterministically complete, without rollback. 

\begin{listing}[t]
\begin{multicols}{2}
\begin{minted}{asm}
    daddu_r_i <-> add_r_i_t {
            $1.reg1 = map_register($$.reg1,64);
            $1.imm = $$.imm;
            $1.dest = map_register($$.dest,64);
            $1.suffix = suf_values[0];
    };
    load_dw <-> i_mov_m_r {
            $1.dest = map_register($$.dest,64);
            $1.offset = $$.offset;
            $1.base = map_register($$.base,64);
            $1.suffix = suf_values[0];
    };
    load_w <-> i_mov_m_r {
            
            $1.dest = map_register($$.dest,32);
            $1.offset = $$.offset;
            $1.base = map_register($$.base,64);
            $1.suffix = suf_values[1];
    };
    store_d <-> i_mov_r_m{
            $1.source = map_register($$.source,64);
            $1.base = map_register($$.base,64);
            $1.offset = $$.offset;
            $1.suffix = suf_values[0];
    };
    store_w <-> i_mov_r_m{
            $1.source = map_register($$.source,32);
            $1.base = map_register($$.base,64);
            $1.offset = $$.offset;
            $1.suffix = suf_values[1];
    };
    addu_r_r <-> add_r_r_t {
            $1.reg1 = map_register($$.reg1,32);
            $1.reg2 = map_register($$.reg2,32);
            $1.dest = map_register($$.dest,32);
            $1.suffix = suf_values[1];
    };
    set_l_r_i <-> set_l_r_i_t {
            $1.reg1 = map_register($$.reg1,32);
            $1.dest = map_register($$.dest,32);
            $1.imm = $$.imm;
            $1.suffix = suf_values[1];
    };
    move <-> i_mov_r_r {
            $1.reg1 = map_register($$.src,64);
            $1.dest = map_register($$.dest,64);
            $1.suffix = suf_values[0];
    };
    balc <-> i_call{
           $1.text = $$.target; 
    };
    branch_c <-> i_jmp{
            $1.text = $$.target;
    };
    branch_eqz_c <-> branch_eq_r_r_t {
            $1.reg1 = map_register($$.reg1,32);
            $1.reg2 = map_register(reg_values[0],32);
            $1.address = $$.target;
            $1.suffix = suf_values[1];
    };
\end{minted}
\end{multicols}
\caption{Example translation part to translate Listing~\ref{lst:fibonacci}.}
\label{lst:example-translation}
\end{listing}

Listing~\ref{lst:example-translation} shows the translation part of the grammar used in our MIPS-to-x86 transformation of the example program in Listing~\ref{lst:fibonacci}. This part of the grammar is used to define one-to-one mappings between MIPS symbols (i.e., abstracted instructions) and the corresponding x86 ones, with accompanying attribute handling that reconstructs the semantics of the source constructs in terms of the target architecture. As can be seen, each rule associates a MIPS non-terminal symbol in the LHS with a corresponding x86 or intermediate symbol on the right, using the \texttt{<->} notation. Attribute values are computed using embedded C code---we recall the inclusion of the \texttt{utils.h} header in the header section of the grammar---e.g. by calling the \texttt{map\_register} function, which performs register translation between the architectures. Also, some instructions behave differently in the two architectures.  ne such example is the \texttt{beq \$t0, \$zero, target\_label} instruction, which branches f the content of the \texttt{\$t0} register is zero. In x86, such an instruction should be mapped to a couple of instructions, like \texttt{cmpl \$0, \%esi}; \texttt{je target\_label}. We include dedicated translation rules (like \texttt{branch\_eqz\_c} for this particular case) to support the translation of one instruction to a group of instructions. The attribute management ensures consistency in the registers\slash constant utilisations.

In addition to translation rules, the grammar supports \textit{reduction} rules, which allow a sequence of source-language symbols to be collapsed into a single target-language symbol. These rules resemble expansion rules in syntax: the LHS is a non-terminal symbol, the RHS consists of one or more symbols, and attribute operations are permitted with \texttt{\$n} and \texttt{\$\$} dereferencing. The arrow symbol is reversed (\texttt{<-}), to indicate abstraction rather than construction. Likewise expansion, reductions may include conditional logic with the \texttt{revert} keyword, enabling the exclusion of invalid derivations based on attribute values.

During generation, each reduction rule is compiled into a function invoked by the parser whenever a new symbol is added. While this allows reductions to be applied incrementally and dynamically, it also introduces computational overhead. Overuse of reduction rules can slow down parsing and generation, so they should be used selectively and only where structural abstraction is semantically justified or it is amenable to reducing the size of the generated\slash translated text.

The integration of expansion, translation, and reduction rules enables the definition of a transformation pipeline that balances flexibility with semantic control. Translation and reduction rules provide deterministic mappings between languages, while expansion rules introduce non-determinism to explore alternative derivations. Relying solely on expansion is insufficient, as it may bypass intended translations and produce incomplete outputs. By combining these mechanisms, the grammar ensures both expressive generation and reliable cross-language transformation.

\subsubsection{Restrictions}

To ensure that the parser generated from the grammar operates deterministically, the grammar must satisfy a set of semantic restrictions that constrain its recognition component, while leaving the generative component non-deterministic (as elaborated in later sections). These restrictions are essential to prevent shift\slash reduce conflicts in the generated parser. Specifically, the following conditions must be satisfied:

\begin{enumerate}[noitemsep]
\item Each non-terminal symbol may be associated with at most one expansion rule whose RHS consists of a single terminal symbol;
\item Each terminal symbol must appear in exactly one expansion rule as an RHS consisting solely of that terminal;
\item Each terminal symbol must have exactly one associated representation rule;
\item All symbol and attribute names defined within the grammar must be globally unique.
\end{enumerate}

The first three constraints ensure parsing determinism. The third condition guarantees that every terminal symbol encountered in the input can be uniquely interpreted according to its representation, while the second ensures that each terminal symbol can be traced back to exactly one non-terminal symbol, allowing unambiguous reconstruction during parsing. Note that this restriction does not limit the expressive capabilities of \framework: in case multiple non-terminal symbols for the same terminal concept are necessary, one can move from one to another by exploiting expansion rules. The first condition prevents representational ambiguity by enforcing a unique derivation path from a non-terminal to its terminal form. The fourth constraint, requiring uniqueness of names across the grammar, is necessary to avoid collisions during C code generation, particularly when symbols are mapped to functions, structures, or global identifiers.

\subsection{The \framework Attribute Grammar Framework}

Now that we have discussed the formal specification of attribute grammars in \framework, we discuss the internal organisation of the framework. As mentioned, \framework allows the generation of code artefacts that can be embedded in working programs to handle the different capabilities of an attribute grammar, namely parsing and (non-deterministically) generating text.

As shown in Figure~\ref{fig:framework-struct}, the \framework framework is composed of two groups of elements: the engine that allows the generation and the static algorithms that are linked to the generated files to form the library. The framework engine consists of a parser for the defined attribute grammar format, built using Flex and Bison, and a generation engine that uses data structure stubs and generation functions to produce the final artefacts.

\subsubsection{Generated Assets}
\label{subsec:library}

\begin{figure}
    \centering
    \includegraphics[width=0.5\linewidth]{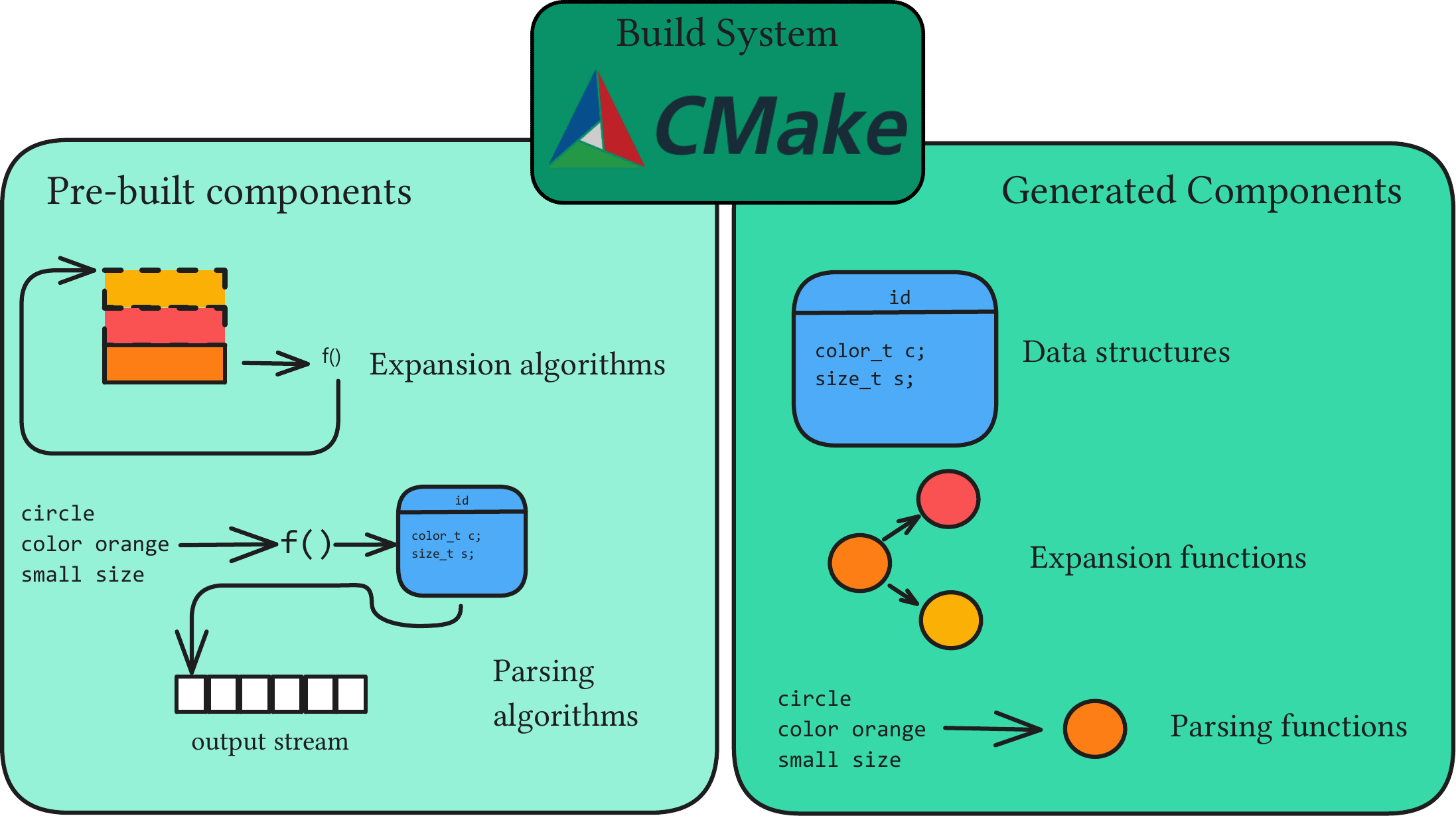}
    \caption{The \framework Framework Organization.}
    \label{fig:library_insights}
\end{figure}

The overall organisation of the \framework framework is depicted in Figure~\ref{fig:library_insights}. We use CMake as the reference build system, which orchestrates the compilation of the grammar, the generation of the C source code from the specification, and the integration of supporting runtime components. 

The input grammar structurally shapes the library generated by the \framework framework, yet it adheres to a well-defined architecture that supports consistent parsing, expansion, and transformation mechanisms.  t a high level, the generated system consists of: (i) a collection of data structures representing the symbols and attributes declared in the grammar, (ii) a suite of functions implementing the user-defined expansion rules, (iii) a deterministic parser tailored to the specific grammar, and (iv) a set of generic data structures and utility routines for grammar-agnostic symbol manipulation.

To enable semantic translation and structural transformation of language constructs, the framework synthesises a dedicated set of data representations that serve as an intermediate layer between the syntactic abstraction of the grammar and its compiled form. These representations preserve the semantics of grammar-defined constructs, while making them amenable to low-level, type-safe manipulation in C.

Each attribute declared in the grammar is mapped to a C type alias that captures its declared type.  or example, an attribute named \texttt{att1} of type \texttt{char *} yields the corresponding definition \texttt{typedef char *att1\_t}. This type aliasing approach facilitates consistency and supports type enforcement across the generated code, particularly in the implementation of expansion and translation rules.

For attributes with a finite domain of enumerated values, the system generates a global array that enumerates all valid instances. The naming convention for these arrays follows the pattern \texttt{attribute\_name\_values}, where \texttt{attribute\_name} is the original grammar-defined name. These arrays provide efficient and direct access to the attribute's semantic domain and support operations such as equality checks, symbolic translation, and randomised generation.

Each symbol defined in the grammar is translated into a C struct, with members corresponding to its declared attributes. The type of each member is determined by the associated attribute's type alias. Every structure is prefixed by an \texttt{unsigned int} field named \texttt{id}, which encodes essential metadata used throughout the parsing and expansion subsystems.

The \texttt{id} field serves a twofold purpose. One bit is reserved to indicate whether the symbol is terminal or non-terminal, while the remaining bits encode a unique numeric identifier assigned at compile time. This compact representation supports efficient symbol discrimination and selection at runtime, allowing expansion and translation functions to operate generically over symbol values without relying on dynamic type information.

A concrete example of the generated data structures, derived from the grammar shown in Listing~\ref{lst:example-declaration}, is shown in Listing~\ref{lst:datastruct}.

\begin{listing}[t]
\begin{minted}{c}
    typedef int num;

    typedef char* reg;

    reg reg_values[] = {"$zero","$v0","$s0","$s1","$sp","$ra"};

    typedef struct __LOAD_DW{
        unsigned int id;
        char *sym_str;
    }LOAD_DW_t;

    typedef struct __load_dw{
        unsigned int id;
        reg dest;
        num offset;
        reg base;
    }load_dw_t;
\end{minted}
\caption{An example of the Data Structures generated starting from the MIPS grammar.}
\label{lst:datastruct}
\end{listing}

\subsubsection{Default Algorithms}

As mentioned, \framework is intended as a grammar workbench, allowing to experiment with different algorithms to manipulate attribute grammars. As such, \framework contains default expansion and parsing algorithms that can be easily replaced or modified---see again the high-level organisation in Figure~\ref{fig:library_insights}. At the same time, these default algorithms allow the \framework user to immediately rely on the generated software artefacts to build additional tools that make use of the expressive power of attribute grammars. In this Section, we discuss the default algorithms shipped with \framework.

\begin{algorithm}[t]
\caption{Grow Algorithm}
\label{algo:grow}
\begin{algorithmic}[1]
\Procedure{\textnormal{grow()}}{}
\State $level\gets 0$
\While{$!is\_empty(stack)$}

\State $sym\gets pop(stack)$

\State $id\gets get\_id(sym)$

\If{$level < max\_level$}
    
       \State $rule\gets extract\_rule(id)$

\Else
    
       \State $rule\gets extract\_terminal\_rule(id)$

\EndIf
    \State $exec\_rule(rule)$

    \State $level\gets level+1$
\EndWhile
\EndProcedure

\vspace{1pt}

\Procedure{\textnormal{extract\_rule}}{unsigned int $sym\_id$}
\State $sym\_rules\gets sym\_to\_rules[sym\_id]$

\State $i\gets 0$
\State $c\gets 0$
\State $rules\gets empty\_array(MAX\_RULES)$

\While{ $i\neq MAX\_RULES \space and \space sym\_rules[i]\geq0$ }

    \If{$bitmap\_check(sym\_rules[i])$}

    \State $i\gets i+1$ 

    \Else
        \State $rules[c]\gets sym\_rules[i]$ 

        \State $c\gets c+1$
        
    \EndIf

\EndWhile

\State $pos\gets random(0,c)$

\State $return \space rules[pos]$
\EndProcedure

\end{algorithmic}
\end{algorithm}

The default expansion mechanism implemented in \framework is the \textit{grow algorithm}, which is inspired by the early work on structured text generation presented in~\cite{Celentano1980-pb}. This algorithm follows a recursive, stack-based strategy for syntactic expansion, where non-terminal symbols are progressively rewritten according to the grammar's rules until a terminal-only sequence is obtained. The complete pseudocode of \framework's implementation of the grow algorithm is presented in Algorithm~\ref{algo:grow}.

As shown in Algorithm~\ref{algo:grow}, the algorithm maintains a generation stack to track symbols that have not yet been expanded. At each step, the symbol at the top of the stack is examined, and a matching expansion or translation rule is selected. A configurable parameter, referred to as the generation depth threshold, can be used to control the maximum recursion depth. When the current depth exceeds or equals this threshold, the rule selection process is constrained to choose only terminal rules---i.e., rules whose RHS consists of a single terminal symbol. This mechanism prevents unbounded recursion and ensures that the generation process eventually converges.

To enable efficient rule selection, each symbol is associated with a unique numeric identifier (\texttt{id}), and a precomputed rule table maps each identifier to the set of applicable rules. This design allows rule lookup to be performed in linear time with respect to the number of rules defined for a given symbol, ensuring both correctness and performance even in grammars with high branching factors.

\lstdefinestyle{customc}{
  language=C,
  basicstyle=\ttfamily\small,
  keywordstyle=\color{blue},
  commentstyle=\color{gray},
  stringstyle=\color{red},
  numbers=left,
  numberstyle=\tiny,
  stepnumber=1,
  numbersep=5pt,
  showstringspaces=false,
  breaklines=true
}
\begin{figure}
    \centering
\begin{tikzpicture}
  % Draw the code box
  \node[anchor=north west, inner sep=0] (code) at (0,0) {
    \begin{tcolorbox}[colback=white, colframe=black, boxrule=0.5pt, sharp corners]
    \begin{lstlisting}[style=customc]
void from circle_to_circle_circle(void *start){


            bitmap_initialize(rule_bitmap,MAX_RULE);
            circle_t *lhs = (circle_t*) start;
            circle_t* circle_0 = malloc(sizeof(circle_t));
            symbol_init(circle_0,"circle",0);
            circle_t* circle_1 = malloc(sizeof(circle_t));
            symbol_init(circle_1,"circle",0);
            
            

            if(lhs->color == color_values[0]){
                   circle_0->color = color_values[1];
                   circle_1->color = color_values[2];
            }else{

                    circle_0->color = lhs->color;
                    circle_1->color = lhs->color;
            } 
            circle_0->size = lhs->size;
            circle_1->size = lhs->size;
            
            
            push(circle_0);
            push(circle_1);
}
    \end{lstlisting}
    \end{tcolorbox}
  };

  % Example arrows and labels
  \node[fill=yellow!30, draw, rounded corners] at (15,-1.05) (label1) {Init};
  \draw[->, thick, yellow] (label1.west) -- ++(-2.5,0) |- ([xshift=150pt,yshift=-40pt]code.north west);

  \node[fill=green!30, draw, rounded corners] at (15,-4) (label2) {Attribute operations};
  \draw[->, thick, green] (label2.west) -- ++(-2.5,0) |- ([xshift=150pt,yshift=-135pt]code.north west);
  \node[fill=orange!30, draw, rounded corners] at (15,-9) (label4) {Finalisation};
  \draw[->, thick, orange] (label4.west) -- ++(-2.5,0) |- ([xshift=150pt,yshift=-270pt]code.north west);
\end{tikzpicture}
    \caption{Generated rule function example}
    \label{fig:rule-generated}
\end{figure}

The structure of a generated expansion rule function is illustrated in Figure~\ref{fig:rule-generated}. Each rule function is composed of three main components: an initialisation phase, an optional attribute evaluation phase, and a finalisation phase. During initialisation, the system restores the global bitmap that encodes the current derivation state and allocates memory for the resulting symbols on the RHS of the rule. If attribute operations are defined for the rule in the grammar, they are executed in the evaluation phase using the dereferencing syntax described previously. In the finalisation phase, the newly constructed symbols are pushed onto the generation stack, preserving the recursive expansion order dictated by the grammar.

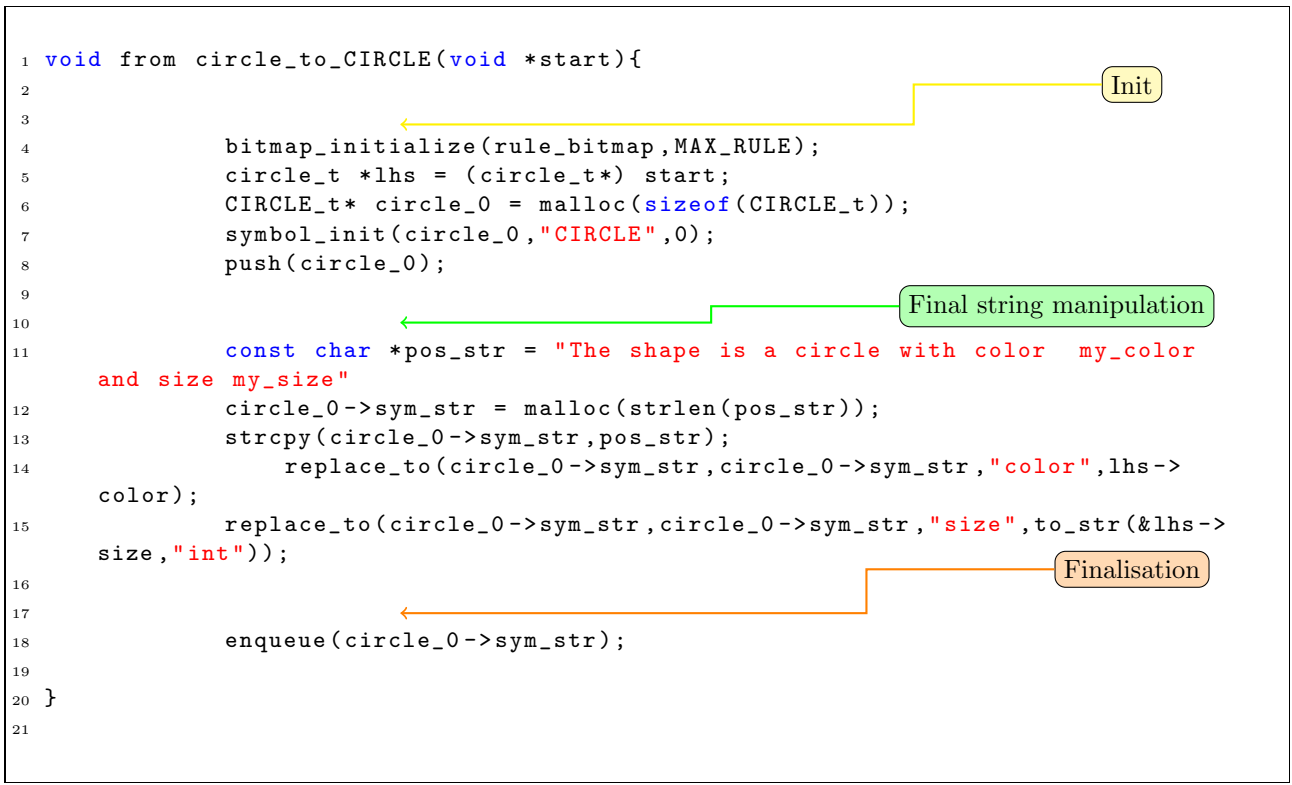
\begin{figure}[t]
    \centering
\begin{tikzpicture}
  % Draw the code box
  \node[anchor=north west, inner sep=0] (code) at (0,0) {
    \begin{tcolorbox}[colback=white, colframe=black, boxrule=0.5pt, sharp corners]
    \begin{lstlisting}[style=customc]
void from circle_to_CIRCLE(void *start){
            
            
            bitmap_initialize(rule_bitmap,MAX_RULE);
            circle_t *lhs = (circle_t*) start;
            CIRCLE_t* circle_0 = malloc(sizeof(CIRCLE_t));
            symbol_init(circle_0,"CIRCLE",0);
            push(circle_0);

            
            const char *pos_str = "The shape is a circle with color  my_color and size my_size"
            circle_0->sym_str = malloc(strlen(pos_str));
            strcpy(circle_0->sym_str,pos_str);
                replace_to(circle_0->sym_str,circle_0->sym_str,"color",lhs->color);
            replace_to(circle_0->sym_str,circle_0->sym_str,"size",to_str(&lhs->size,"int"));
            
            
            enqueue(circle_0->sym_str);

}
    \end{lstlisting}
    \end{tcolorbox}
  };

  % Example arrows and labels
  \node[fill=yellow!30, draw, rounded corners] at (15,-1.05) (label1) {Init};
  \draw[->, thick, yellow] (label1.west) -- ++(-2.5,0) |- ([xshift=150pt,yshift=-45pt]code.north west);

  \node[fill=green!30, draw, rounded corners] at (14,-4) (label2) {Final string manipulation};
  \draw[->, thick, green] (label2.west) -- ++(-2.5,0) |- ([xshift=150pt,yshift=-120pt]code.north west);
  \node[fill=orange!30, draw, rounded corners] at (15,-7.5) (label4) {Finalisation};
  \draw[->, thick, orange] (label4.west) -- ++(-2.5,0) |- ([xshift=150pt,yshift=-230pt]code.north west);
\end{tikzpicture}
\caption{Generated terminal rule function example}
    \label{fig:terminal-generated}
\end{figure}

Terminal rules follow a similar structural pattern but are tailored to produce text rather than further symbolic derivations. The general form of a generated terminal rule is depicted in Figure~\ref{fig:terminal-generated}. As in non-terminal rules, the function begins with an initialisation phase, during which attributes are evaluated and the symbol's state is prepared. This is followed by a substitution phase, where the terminal representation string is assembled by replacing attribute placeholders with actual values. The result is then appended to a queue of generated strings. At the end of the ``grow'' process, this queue is concatenated to form the final output.

\begin{figure}[t]
    \centering
    \includegraphics[width=0.6\linewidth]{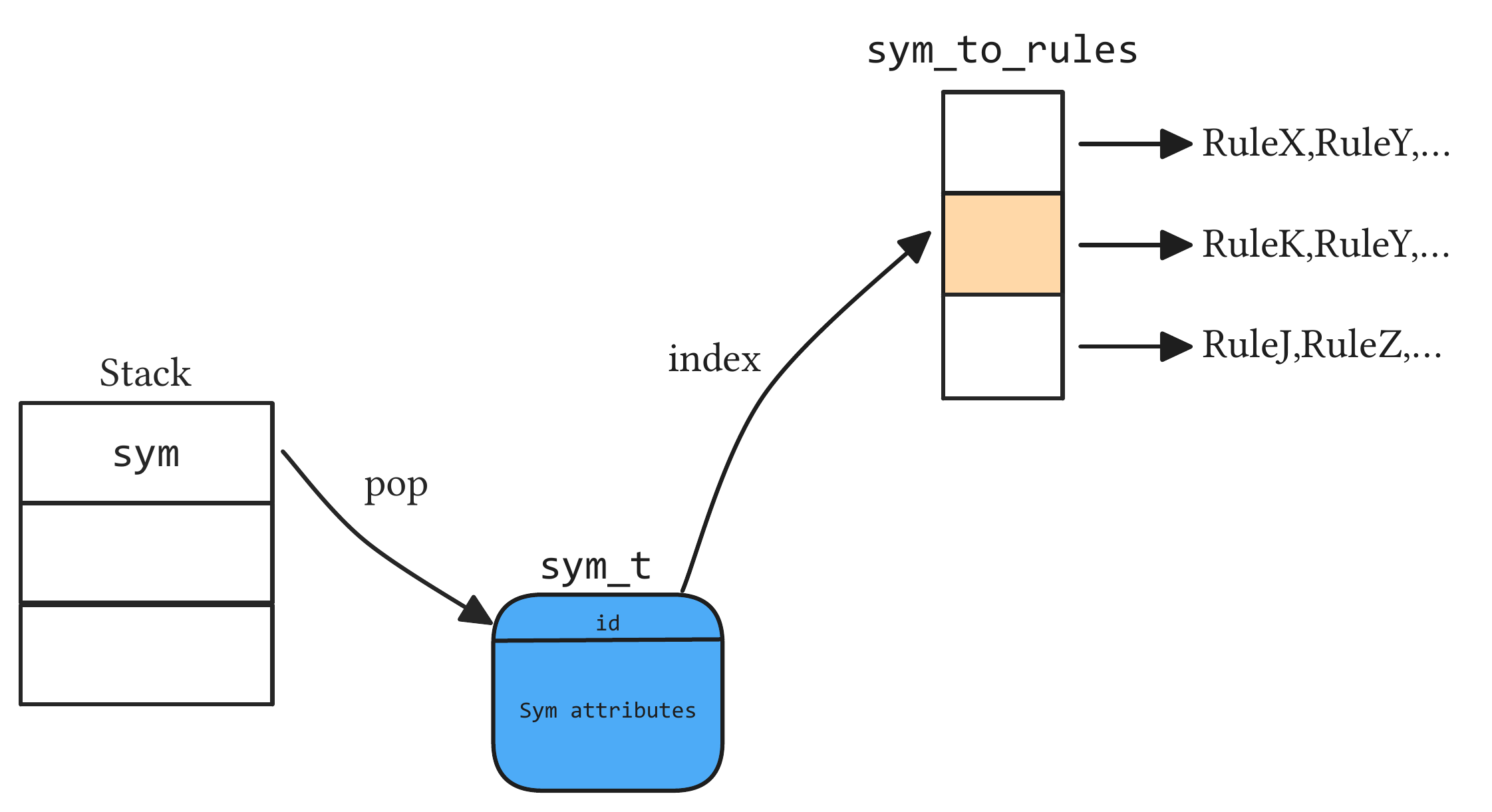}
    \caption{The Grow Algorithm Operations.}
    \label{fig:grow-alg}
\end{figure}

Conversely, the default parsing mechanism implemented in \framework is built upon Flex and Bison, which are used to perform lexical and syntactic analysis, respectively. These tools generate parsers that interact with a group of \framework's initialisation routines and transformation functions, which allocate data structures and write the parsed output to a preallocated memory region. This mechanism bridges the textual representation of the language with its internal attribute representation.

The parser exposes a unified interface centred on a single entry point, the \texttt{parse} function. The presence of this single entry point can be exploited for algorithm replacement if \framework shall be used as an algorithmic workbench. Before invoking this function, both the input and output streams must be configured. The input stream can be either a character buffer or a filename containing the source data to be parsed. The output stream, on the other hand, must be a pointer to a contiguous memory region sufficiently large to accommodate the data structures generated during parsing. This memory is populated with one structure per symbol parsed, though this upper bound may be reduced due to the application of reduction rules.

Reductions are applied dynamically during parsing to compact sequences of symbols into higher-level constructs as defined by the grammar. Each reduction rule is triggered when a contiguous subsequence of symbols in the output stream matches the rule's RHS. After each symbol insertion, the system evaluates all applicable reduction conditions. If a match is found, the associated reduction function is executed, replacing the matched sequence with a single symbol that encapsulates its aggregated semantics. These functions are automatically generated from the grammar, including any user-defined attribute operations.

Although powerful, this mechanism introduces computational overhead: the cost of evaluating reduction triggers grows with the number and complexity of reduction rules. Excessive reliance on reductions may thus degrade parsing performance. A high-level view of the parser control flow is provided in Algorithm~\ref{algo:parse}.

The Flex and Bison grammars required to support this process are also generated by \framework. The generated lexer rule section includes regular expressions for all attributes defined over finite domains, along with patterns that match quoted strings in terminal representation rules. At the end of the lexer file, regular expressions for attributes with infinite domains are included---either explicitly defined by the user or based on default patterns if no definition is given. These expressions are sorted first by their order in the grammar file, and if multiple grammar files are used, by the order of the files and then the order within each file.

The generated parser file defines a token for each lexeme recognised by the lexer, along with a grammar rule for each terminal symbol, as derived from its representation rule. Each rule is paired with a semantic action that constructs the corresponding data structure and stores it in the output stream. These actions directly instantiate the types generated from the grammar and initialise their attributes based on the matched values. 

\begin{algorithm}[!t]
\caption{Parse Algorithm}
\label{algo:parse}
\begin{algorithmic}[1]
\Procedure{\textnormal{parse()}}{}
\State $input\gets input\_stream$
\State $flex\_and\_Bison(input);$
\EndProcedure

\vspace{1pt}

\Procedure{\textnormal{save\_symbol}}{void* $sym$}
\If{$n\_symbols < size$}
    \State $output\_stream[n\_symbols]\gets sym$
    \State $n\_symbols\gets n\_symbols+1$
\Else
    \State $end\_parse();$
\EndIf
\EndProcedure

\Procedure{\textnormal{reduce\_check}}{}
\State $prev\_symbols[MAX]\gets retrieve(MAX)$
\State $i\gets 0$
\While{$i < n\_reduction\_rules$}
    \If{$condition\_met(prev\_symbols,i)$}
        \State $reductions[i](prev\_symbols);$
    \EndIf
    \State $i\gets i+1$
\EndWhile
\EndProcedure

\end{algorithmic}
\end{algorithm}

\section{Experimental Assessment}
\label{sec:experimental}

In this Section, we present an experimental evaluation of \framework to verify the \textit{correctness} and \textit{applicability} of the software artefacts it generates. The primary objective of this assessment is not performance benchmarking, but rather to validate that the translations produced by \framework are semantically consistent and operationally sound across realistic, non-trivial programs.

To evaluate \framework under more challenging conditions than the Fibonacci example of Listing~\ref{lst:fibonacci}, which served only as an illustrative, pedagogical case to expose the main design principles and features of \framework, we apply the MIPS-to-x86 translational grammar discussed above to a subset of the MiBench benchmark suite~\cite{Guthaus2005-hj}. MiBench is a collection of embedded-oriented programs designed to reflect real-world usage scenarios.

We selected three representative benchmarks from MiBench: \textit{Qsort}, \textit{Dijkstra}, and \textit{Stringsearch}. These programs are sufficiently complex to test critical translation aspects such as register allocation, memory access patterns, adherence to the target calling conventions, and the correct propagation of control flow. Specifically, `Qsort` involves function pointers and system-level library calls, `Dijkstra` stresses pointer arithmetic and iterative control structures over 2D data structures, and `Stringsearch` exercises low-level character manipulation and table-based logic.

To illustrate the versatility of \framework, we have also applied a second \framework-generated translation to the intermediate representation used by the \textsc{VMorph} virtualisation platform~\cite{Caliandro2025-xn}. To support both translations, we constructed dedicated grammar specifications for the x86 and \textsc{VMorph} targets and reused the existing MIPS grammar. This approach highlights \framework's ability to extend and interconnect multiple language specifications via translation rules without rewriting the entire grammars.

Each MiBench benchmark was first compiled to the MIPS instruction set using a cross-compiler configured for the MIPS64 Release 6 ABI. The resulting binaries were disassembled to extract MIPS assembly code, which was then translated using the software assets generated with \framework into the two distinct target languages. 

Given the complexity of the MIPS architecture, we limited our support to a semantically complete subset sufficient to express all constructs used in the selected benchmarks. To ensure conformance with this subset, the MIPS compiler was invoked with a curated set of flags: \texttt{-march=mips64r6}, \texttt{-mabi=64}, and \texttt{-mno-abicalls} restrict code generation to the supported instruction set; \texttt{-fno-pic} and \texttt{-static} disable position-independent code; and \texttt{-fno-stack-protector} eliminates stack-related instrumentation irrelevant for translation. For \textsc{VMorph}, we additionally pass \texttt{-fomit-frame-pointer} due to the lack of a dedicated frame pointer register in the virtual ISA.

Each translated program was manually executed, and its output was compared against the output of the original MIPS binary. In every case, the outputs were found to be equivalent, demonstrating the correctness of the transformation pipeline from parsing to final code generation. 

To support reproducibility and allow independent verification of these results, we include in the supplemental material (see the persistent version of the code available at \url{https://doi.org/10.5281/zenodo.15357268}) the full sources of the programs, as well as the grammar files, instructions to re-run the experiments, and any necessary driver code. This material enables the reader to inspect the entire pipeline and validate the functional equivalence of the input and translated programs directly.

\section{Conclusions}

In this paper, we introduced \framework, an attribute grammar framework system for the specification and automated generation of software artefacts able to parse, generate and translate text based on an attribute grammar specification. \framework can be regarded as an algorithmic workbench, as it allows the replacement of its default expansion\slash parsing algorithms easily.  ur design supports both deterministic and non-deterministic attribute grammars.
We demonstrated the applicability of \framework on real-world benchmarks from the MiBench suite, showing that it can correctly translate MIPS programs into both x86 and virtualised \textsc{VMorph} code. All generated outputs were manually verified to be functionally equivalent to their source programs.

\section*{Acknowledgements}

The authors are grateful to Prof. Alberto Pettorossi for his invaluable feedback on the initial ideas and early versions of this work.

\bibliographystyle{acm}
\bibliography{library,matteo}%

\begin{thebibliography}{10}

\bibitem{Aho2007-kw}
{\sc Aho, A.~V., Sethi, R., Ullman, J.~D., Lam, M.~S., Sethi, R., and Ullman, J.~D.}
\newblock {\em Compilers: Principles, Techniques, and Tools}, 2~ed.
\newblock Prentice Hall, 2007.

\bibitem{Augusto2023-sc}
{\sc Augusto, L.~M.}
\newblock The van wijngaarden grammars: A syntax primer with decidable restrictions.
\newblock {\em Journal of Knowledge Structures and Systems 4}, 2 (2023), 1--39.

\bibitem{Baker2018-da}
{\sc Baker, M.}
\newblock {\em In other words: A coursebook on translation}, 3~ed.
\newblock Routledge, Apr. 2018.

\bibitem{Bisbal1999-ay}
{\sc Bisbal, J., Lawless, D., Wu, B., and Grimson, J.}
\newblock Legacy information systems: issues and directions.
\newblock {\em IEEE software 16}, 5 (1999), 103--111.

\bibitem{Boyland2005-gn}
{\sc Boyland, J.~T.}
\newblock Remote attribute grammars.
\newblock {\em Journal of the ACM 52}, 4 (July 2005), 627--687.

\bibitem{Caliandro2025-xn}
{\sc Caliandro, P., Ciccaglione, M., Pepe, A., Bianchi, G., and Pellegrini, A.}
\newblock {VMORPH}: A virtualization/metamorphic framework for binary obfuscation and intellectual property protection.
\newblock In {\em Proceedings of the 2025 Italian Conference on Cybersecurity\/} (Bologna, Italy, Feb. 2025), ITASEC, CEUR-WS.org.

\bibitem{Celentano1980-qe}
{\sc Celentano, A., Reghizzi, S.~C., Vigna, P.~D., Ghezzi, C., Granata, G., and Savoretti, F.}
\newblock Compiler testing using a sentence generator.
\newblock {\em Software: practice \& experience 10}, 11 (Nov. 1980), 897--918.

\bibitem{Celentano1980-pb}
{\sc Celentano, A., Reghizzi, S.~C., Vigna, P.~D., Ghezzi, C., Granata, G., and Savoretti, F.}
\newblock Compiler testing using a sentence generator.
\newblock {\em Softw. Pract. Exp. 10}, 11 (Nov. 1980), 897--918.

\bibitem{Chen2021-ax}
{\sc Chen, M., Tworek, J., Jun, H., Yuan, Q., Pinto, H. P. d.~O., Kaplan, J., Edwards, H., Burda, Y., Joseph, N., Brockman, G., Ray, A., Puri, R., Krueger, G., Petrov, M., Khlaaf, H., Sastry, G., Mishkin, P., Chan, B., Gray, S., Ryder, N., Pavlov, M., Power, A., Kaiser, L., Bavarian, M., Winter, C., Tillet, P., Such, F.~P., Cummings, D., Plappert, M., Chantzis, F., Barnes, E., Herbert-Voss, A., Guss, W.~H., Nichol, A., Paino, A., Tezak, N., Tang, J., Babuschkin, I., Balaji, S., Jain, S., Saunders, W., Hesse, C., Carr, A.~N., Leike, J., Achiam, J., Misra, V., Morikawa, E., Radford, A., Knight, M., Brundage, M., Murati, M., Mayer, K., Welinder, P., McGrew, B., Amodei, D., McCandlish, S., Sutskever, I., and Zaremba, W.}
\newblock Evaluating large language models trained on code.
\newblock {\em arXiv [cs.LG]\/} (July 2021).

\bibitem{Chomsky1965-nf}
{\sc Chomsky, N.}
\newblock {\em Aspects of the theory of syntax}, 1~ed.
\newblock The MIT Press. MIT Press, London, England, 1965.

\bibitem{Chomsky2019-nh}
{\sc Chomsky, N.}
\newblock {\em Studies on semantics in generative grammar}, 3~ed.
\newblock Janua Linguarum. Series Minor. De Gruyter Mouton, Berlin, Germany, May 2019.

\bibitem{Devereaux2010-ki}
{\sc Devereaux, J.~E.}
\newblock Obsolescence: A systems engineering and management approach for complex systems.
\newblock Master's thesis, Massachusetts Institute of Technology, Cambridge, MA, USA, Feb. 2010.

\bibitem{Donnelly2015-rr}
{\sc Donnelly, C., and Stallman, R.}
\newblock {\em Bison: The Yacc-compatible parser generator}.
\newblock Samurai Media Ltd, Guelph, ON, USA, Nov. 2015.

\bibitem{Erdweg2012-vf}
{\sc Erdweg, S., Giarrusso, P.~G., and Rendel, T.}
\newblock Language composition untangled.
\newblock In {\em Proceedings of the Twelfth Workshop on Language Descriptions, Tools, and Applications\/} (New York, NY, USA, Mar. 2012), ACM.

\bibitem{Ghete2018-ni}
{\sc Ghete, M.~C.}
\newblock Extending bison with attribute grammars.
\newblock Master's thesis, Technische Universität Wien, Wien, 2018.

\bibitem{Guthaus2005-hj}
{\sc Guthaus, M.~R., Ringenberg, J.~S., Ernst, D., Austin, T.~M., Mudge, T., and Brown, R.~B.}
\newblock {MiBench}: A free, commercially representative embedded benchmark suite.
\newblock In {\em Proceedings of the Fourth Annual IEEE International Workshop on Workload Characterization. WWC-4 (Cat. No.01EX538)\/} (2005), IEEE.

\bibitem{Hedin2011-oy}
{\sc Hedin, G.}
\newblock An introductory tutorial on {JastAdd} attribute grammars.
\newblock In {\em Lecture Notes in Computer Science}, Lecture notes in computer science. Springer Berlin Heidelberg, Berlin, Heidelberg, 2011, pp.~166--200.

\bibitem{Hutchins1992-bu}
{\sc Hutchins, W.~J., and Somers, H.~L.}
\newblock {\em An introduction to machine translation}.
\newblock Academic Press, 1992.

\bibitem{Knuth1968-ed}
{\sc Knuth, D.~E.}
\newblock Semantics of context-free languages.
\newblock {\em Mathematical Systems Theory 2}, 2 (June 1968), 127--145.

\bibitem{Levine2009-dl}
{\sc Levine, J.}
\newblock {\em flex \& bison}.
\newblock O'Reilly Media, Aug. 2009.

\bibitem{Li2022-sx}
{\sc Li, Y., Choi, D., Chung, J., Kushman, N., Schrittwieser, J., Leblond, R., Eccles, T., Keeling, J., Gimeno, F., Dal~Lago, A., Hubert, T., Choy, P., de~Masson~d'Autume, C., Babuschkin, I., Chen, X., Huang, P.-S., Welbl, J., Gowal, S., Cherepanov, A., Molloy, J., Mankowitz, D.~J., Sutherland~Robson, E., Kohli, P., de~Freitas, N., Kavukcuoglu, K., and Vinyals, O.}
\newblock Competition-level code generation with {AlphaCode}.
\newblock {\em Science (New York, N.Y.) 378}, 6624 (Dec. 2022), 1092--1097.

\bibitem{Moonen2001-wt}
{\sc Moonen, L.}
\newblock Generating robust parsers using island grammars.
\newblock In {\em Proceedings Eighth Working Conference on Reverse Engineering\/} (Piscataway, NJ, USA, 2001), IEEE Comput. Soc, pp.~13--22.

\bibitem{Paakki1995-pp}
{\sc Paakki, J.}
\newblock Attribute grammar paradigms—a high-level methodology in language implementation.
\newblock {\em ACM computing surveys 27}, 2 (June 1995), 196--255.

\bibitem{Parr1995-vj}
{\sc Parr, T.~J., and Quong, R.~W.}
\newblock {ANTLR}: A predicated‐\textit{LL}(\textit{k}) parser generator.
\newblock {\em Software: practice \& experience 25}, 7 (July 1995), 789--810.

\bibitem{Paxson2023-ey}
{\sc Paxson, V., Estes, W., and Millaway, J.}
\newblock Lexical analysis with flex, Jan. 2023.

\bibitem{Purdom1972-fl}
{\sc Purdom, P.}
\newblock A sentence generator for testing parsers.
\newblock {\em BIT numerical mathematics 12}, 3 (Sept. 1972), 366--375.

\bibitem{Seacord2003-bw}
{\sc Seacord, R.~C., Plakosh, D., and Lewis, G.~A.}
\newblock {\em Modernizing legacy systems: software technologies, engineering processes, and business practices}.
\newblock Addison-Wesley Professional, Boston, MA, USA, 2003.

\bibitem{van-Wijngaarden1965-sp}
{\sc van Wijngaarden, A.}
\newblock Orthogonal design and description of a formal language.
\newblock Tech. Rep. MR 76, Mathematisch Centrum, Amsterdam, NL, 1965.

\bibitem{Van-Wyk2010-vw}
{\sc Van~Wyk, E., Bodin, D., Gao, J., and Krishnan, L.}
\newblock Silver: An extensible attribute grammar system.
\newblock {\em Science of computer programming 75}, 1-2 (Jan. 2010), 39--54.

\bibitem{Van-Wyk2007-rt}
{\sc Van~Wyk, E., Krishnan, L., Bodin, D., and Schwerdfeger, A.}
\newblock Attribute grammar-based language extensions for java.
\newblock In {\em ECOOP 2007 – Object-Oriented Programming}, Lecture notes in computer science. Springer Berlin Heidelberg, Berlin, Heidelberg, 2007, pp.~575--599.

\bibitem{Wang2021-oh}
{\sc Wang, Y., Wang, W., Joty, S., and Hoi, S. C.~H.}
\newblock {CodeT5}: Identifier-aware unified pre-trained encoder-decoder models for code understanding and generation.
\newblock {\em arXiv [cs.CL]\/} (Sept. 2021).

\end{thebibliography}

\end{document}